\documentclass[amsmath,amssymb,prc,final,showpacs,twocolumn]{revtex4-1}
 
\usepackage{bm,hyphenat,xspace}
\usepackage{graphicx,epsfig}

\newcommand {\anz}{a_n^-}

\newcommand{\etal}{{\em et al.}}

\begin{document}

\title {Energies and widths of Efimov states in the three-boson continuum}
  
\author{A.~Deltuva} 
\email{arnoldas.deltuva@tfai.vu.lt}
\affiliation
{Institute of Theoretical Physics and Astronomy, 
Vilnius University, Saul\.etekio al. 3, LT-10257 Vilnius, Lithuania
}

\received{April 21, 2020} 

\begin{abstract}
Three-boson Efimov physics is well known in the bound-state regime, but 
far less in the three-particle continuum at negative two-particle scattering length
where Efimov states evolve into resonances. They are studied solving
rigorous three-particle scattering equations for transition operators
in the momentum space. The dependence of the three-boson resonance energy and width
on the two-boson scattering length is studied with several force models.
The universal limit is determined numerically considering highly excited states;
simple parametrizations for the resonance energy and width in terms of the scattering length
are established. Decreasing the attraction, the resonances rise not much above the threshold but broaden rapidly 
and become physically unobservable, evolving into subthreshold resonances.
Finite-range effects are studied and related to those in the bound-state regime.
\end{abstract}

 \maketitle

\section{Introduction \label{sec:intro}}

Fifty years ago V. Efimov studied theoretically the three-body system with large
two-body scattering lengths \cite{efimov:plb}
and laid the foundations
of the universal physics, also called Efimov physics.
Since then a large number of theoretical and experimental works with applications
to nuclear, cold atom, and molecular physics has been performed, 
and the properties of universal few-body systems have been determined;
see recent reviews \cite{naidon:rev,greene:rev} for summary and further references. 
In particular, the bound-state energy levels of three identical bosons 
and their dependence on the two-boson scattering length is among the best-known manifestations
of the Efimov physics, even with semi-analytic parametrizations
available  \cite{naidon:rev,braaten:rev},  the most refined version being presented
in Ref.~\cite{gattobigio:19a}.
However,  the universal properties of the three-boson system 
were studied less extensively in the regime of negative scattering length 
above the free-particle threshold where the trimers  become unbound.
 Reference \cite{braaten:rev} considered only the limit of vanishing energy,
while Refs.~\cite{braaten:09a,Wang_2011,jonsell:06a} investigated the recombination into deeply bound dimers.
It is known that in this regime the trimers become resonant states \cite{bringas:04a};
their energies, widths,
and their evolution with the scattering length were determined by Bringas \etal{} \cite{bringas:04a},
but not  in a strictly universal regime where the scattering length is sufficiently large
such that the finite-range effects are negligible. A systematic study of finite-range effects
is also still missing.
The aim of the present work is to resolve this situation and to
present a high-precision study of universal three-boson resonant states
in the three-particle continuum with no bound dimers, neither shallow
nor deep.

The study is based on exact Faddeev-type equations \cite{faddeev:60a} for three-body transition
operators in the version proposed by Alt, Grassberger, and Sandhas (AGS)
\cite{alt:67a}, 
that are solved in the momentum space. The method has been used for
the search for three-neutron resonances \cite{deltuva:18a}.
An important advantage 
of the direct continuum approach is its ability to estimate not only the 
resonance position but also its effect on scattering amplitudes
that lead to observables in collision processes.
Additional complications in the context of Efimov physics arise due to very different sizes
of the interaction range and the characteristic interparticle distances.

Section \ref{sec:eq} describes three-particle scattering equations
and some details of calculations whereas Sec. III contains
results. The summary is presented in Sec. IV.

\section{Theory \label{sec:eq}}

If the two-particle potential $v$ is tuned to yield a 
negative two-boson scattering length $a$ and no bound dimers, then
the only possible process in the three-boson continuum is the
$3\to 3$  scattering. As given in Ref.~\cite{deltuva:18a},
the symmetrized transition operator for this process,
\begin{equation} \label{eq:AGS00s}
U_{00} =  (1+P)t(1+P)  + (1+P) t G_0 U G_0 t (1+P)
\end{equation}
contains two-body  and three-body parts represented by
the first and second terms, respectively;
 only the latter  corresponds to a genuine three-particle process. 
All operators in Eq.~(\ref{eq:AGS00s}) refer to the
relative three-particle motion. They are the 
combination of particle permutation operators
$P =  P_{12}\, P_{23} + P_{13}\, P_{23}$,
the free resolvent 
\begin{equation} \label{eq:G0}
G_0 = (E+i0-H_0)^{-1}
\end{equation}
at the available energy $E$ with the kinetic energy operator $H_0$,
 and the two- and three-particle
transition operators $t$ and $U$, respectively.
 They are related to the two-particle potential $v$ via integral equations
by Lippmann-Schwinger,
\begin{equation} \label{eq:t}
t = v  + v G_0 t 
\end{equation}
and AGS 
\begin{equation} \label{eq:AGSs}
U =  P G_0^{-1} + P t\, G_0 \, U.
\end{equation}

As  is well known, a true bound state 
of the few-body system (stable with respect to decay into several clusters)
corresponds to the pole of the respective transition operator
at a real negative energy $E= - E_b$ in the physical energy sheet,
$E_b$ being the binding energy.  In the system of three spinless bosons,
bound Efimov states have zero total angular momentum $J=0$ and positive parity $\Pi$.
When this bound state, via variation of the interaction, crosses the 
free-particle threshold and becomes a resonance, the pole of the
transition operators (both $U$ and $U_{00}$) moves to $E = E^R -i \Gamma/2$ in the 
nonphysical complex energy sheet, $E^R$ being the resonance position 
and $\Gamma$ its width. 
As long as $E^R> 0$  and $\Gamma$ is sufficiently small,
the pole is not far from the physical scattering region and manifests itself
via resonant enhancement of continuum processes. This is reflected
in the energy dependence of the corresponding transition
operator, that can be expanded into power series near the pole, i.e.,
\begin{equation} \label{eq:Upole}
 U_{J^\Pi} =  \sum_{n_r=-1}^\infty \tilde{U}^{(n_r)}_{J^\Pi} \, (E-E^R + i \Gamma/2 )^{n_r},
\end{equation}
the $n_r=-1$ term being the pole term and $n_r \ge 0$ being background terms.
In this way one can  determine not only the position and width of the resonance,
but also its importance relative to the non-resonant background.
To a very good approximation, in the vicinity of the pole the series (\ref{eq:Upole})
can be truncated  after few lowest terms while
higher-order terms yield negligible contribution.

The procedure of determining the resonance parameters follows closely
the one of Ref.~\cite{deltuva:18a} for three-neutron resonances.
The AGS equation (\ref{eq:AGSs}) is solved in the momentum-space 
partial wave representation; for technical details,  including also the treatment
of kinematical singularities in the kernel of the equation, see also
Ref.~\cite{deltuva:phd}. 
Various on-shell and off-shell matrix elements of three-boson transition
operators $U$ and $T=t G_0 U G_0 t $ are calculated as functions of the energy $E$,
within numerical accuracy all exhibiting the same resonant behavior,
much like that shown in Ref.~\cite{deltuva:18a} for the three-neutron
system with enhanced interaction. 
The  position and width of the resonance is determined fitting 
those energy-dependent matrix elements to Eq.~(\ref{eq:Upole}).
Differences between the resonance parameters  obtained from 
transition matrix elements with different initial and final test states,
as well as variations due to the different maximal power $n_r$ 
in the expansion (\ref{eq:Upole}), serve as an estimation of the error for this procedure.
Typically,  for small $\Gamma$ the pole is close to the physical scattering region,
the resonant behavior is pronounced very clearly, and, consequently, 
the error is extremely small. In contrast, very broad resonances become hardly 
distinguishable from the background, resulting also slower convergence of
the expansion (\ref{eq:Upole}) where $n_r$ up to 5 or 6 was employed in the present
work. This leads to decreasing accuracy, and renders a reliable extraction of
the resonance parameters impossible when $\Gamma$ is too large.
On the other hand, the fact that resonant behavior cannot be seen in transition
matrix elements indicates that the resonance cannot be distinguished 
from background and becomes physically unobservable.

\section{Results \label{sec:res}}

The present work aims to study the evolution of three-boson Efimov resonances,
i.e., to establish universal relations between their energies and widths
and the two-boson scattering length, similar to what is well known
for bound states \cite{naidon:rev,braaten:rev,gattobigio:19a}.
The quantity connecting the trimer bound- and resonant-state regimes
is the negative two-boson scattering length $\anz$ where the
$n$-th Efimov trimer crosses the free-particle threshold.
There is a universal relation between $\anz$
and the binding momentum $k_n^u$ of the $n$-th Efimov trimer
in the unitary limit. Reference \cite{braaten:rev}, where
also the elastic three-boson scattering in the zero energy limit was considered,
predicted $k_n^u \anz = -1.56(5)$. However, the most accurate
results are $k_n^u \anz = -1.50763$,
obtained in Ref.~\cite{PhysRevLett.100.140404} analytically, and
$k_n^u \anz = -1.5077(1)$, obtained numerically in 
Ref.~\cite{deltuva:12a} considering highly excited Efimov states  up to $n=5$.
To make the connection with the latter work, the two-boson interaction
model is taken from Ref.~\cite{deltuva:12a}. It has only the $S$-wave
component
\begin{gather} \label{eq:lsep}
\begin{split}
  \langle p_f | v_s | p_i \rangle = & (1+ \beta  p_f^2/\Lambda^2 ) \,  e^{-p_f^2/\Lambda^2} \\ &
\times \frac{2}{\pi m}
\left\{ \frac1a - \left[1+\frac{\beta}{2} \left(1+\frac{3\beta}{8} \right) \right] 
\frac{\Lambda}{\sqrt{2\pi}} \right\}^{-1} \\ &
\times 
e^{-p_i^2/\Lambda^2} \, (1+ \beta  p_i^2/\Lambda^2 )
\end{split}
\end{gather}
where $p_f$ ($p_i$) is the final (initial) momentum, $m$ is the boson mass,
and $\Lambda$ is the momentum cutoff parameter that controls the range of the force;
the $\hbar=1$ convention is used. The parameter $\beta$ enables variations in the balance 
of low- and  high-momentum components as will be needed later on; for the following results
it is, however, chosen as $\beta =0$  unless explicitly stated otherwise.
For the extraction of  universal relations highly excited states
have to be considered with vanishing finite-range corrections.
This can be characterized by the ratio of the two-boson $S$-wave effective range
$r_s$ and the scattering length $a$. At the trimer and free-particle threshold
intersection points  these ratios $r_s/|\anz|$ are
$0.372, \, 0.0187, \, 8.45\times 10^{-4}$, and  $3.73\times 10^{-5}$
for $n=0$, 1, 2, and 3, respectively. The trimer ground state $n=0$ is 
therefore expected to exhibit significant finite-range corrections. However,
finite-range effects should become very small for $n \ge 2$. Indeed,
the product $k_n^u \anz$ is $-2.0629, \, -1.5642, \, -1.5116$, and 
$-1.5079$ for $n=0$, 1, 2, and 3, respectively,
in the $n=3$ case being very close to the universal value of
$-1.50763$. This suggests that  $n=3$ should be sufficient also
for the determination of universal Efimov resonance properties
with a good accuracy.

\begin{figure}[!]
\begin{center}
\includegraphics[scale=0.66]{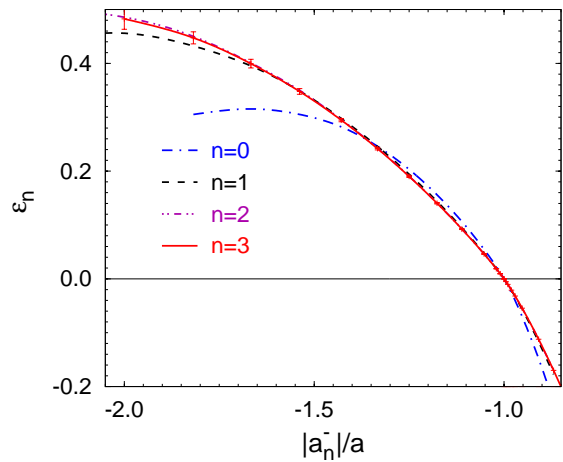}
\end{center}
\caption{\label{fig:Er0-3} (Color online)
Dimensionless real energy part
for ground ($n=0$) and excited ($n=1$, 2, and 3) 
three-boson states as a function of the inverse two-boson scattering length.
}
\end{figure}

\begin{figure}[!]
\begin{center}
\includegraphics[scale=0.64]{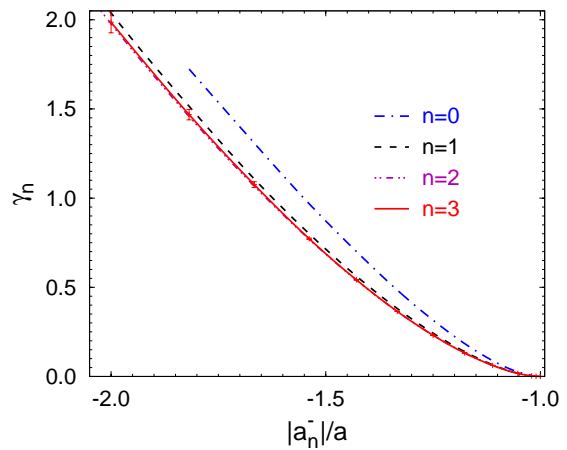}
\end{center}
\caption{\label{fig:G0-3} (Color online)
Dimensionless width
for ground ($n=0$) and excited ($n=1$, 2, and 3) 
three-boson states as function of the inverse two-boson scattering length.
 }
\end{figure}

To emphasize the universal character of the results, they will be given as dimensionless
quantities. Thus, the energy $E^R_n$ and width $\Gamma_n$ of the $n$-th Efimov state 
will be shown in the dimensionless forms
$\varepsilon_n = E^R_n \, m (\anz)^2/\hbar^2 $
and $\gamma_n =  \Gamma_n \, m (\anz)^2/\hbar^2$.
In other words, $\varepsilon_n$ ($\gamma_n $) is the energy (width)
of the  $n$-th Efimov resonance in units of $\hbar^2/[m (\anz)^2]$.
For the two-boson scattering length the most meaningful reference point in the present
context is $\anz$  for each $n$. 
To keep consistency with the standard representation of the Efimov 
physics in terms of the inverse scattering length, the results will be
shown as functions of the dimensionless quantity
$|\anz|/a$ that takes values below (above) $-1$ in the resonance (bound state) regime.
The dimensionless energies $\varepsilon_n$ 
of four lowest three-boson states with $n=0,\, 1,\, 2$, and  3
are shown in Fig.~\ref{fig:Er0-3},
 partially including also the bound state region 
where the binding energy was obtained by solving the standard bound-state Faddeev equation
\cite{deltuva:12a}. The bound-state and continuum calculation results connect
well at $|\anz|/a = -1$. 
The corresponding dimensionless resonance widths $\gamma_n$ are  shown in Fig.~\ref{fig:G0-3}.
In both cases the convergence with increasing $n$ is evident, suggesting that,
$n=2$ and $n=3$ results, being very close to each other, accurately
represent the universal limit. In contrast, the ground state ($n=0$) results
show significant deviations  from the universal limit  due to finite-range corrections,
as can be expected given the corresponding ratio
$r_0/|\anz| = 0.372$.
The width of  the resonance increases with decreasing
$|\anz|/a$, and, as a consequence, the theoretical error bars increase too.
For clarity, the theoretical error bars are given only for 
$n=3$ results, they are roughly the same for all considered $n$ values.
Beyond $|\anz|/a = -2$ a reliable extraction of resonance properties 
is very difficult. The resonant behavior becomes very broad and unobservable
since the width exceeds the energy by a factor of 4 or more.

\begin{figure}[!]
\begin{center}
\includegraphics[scale=0.64]{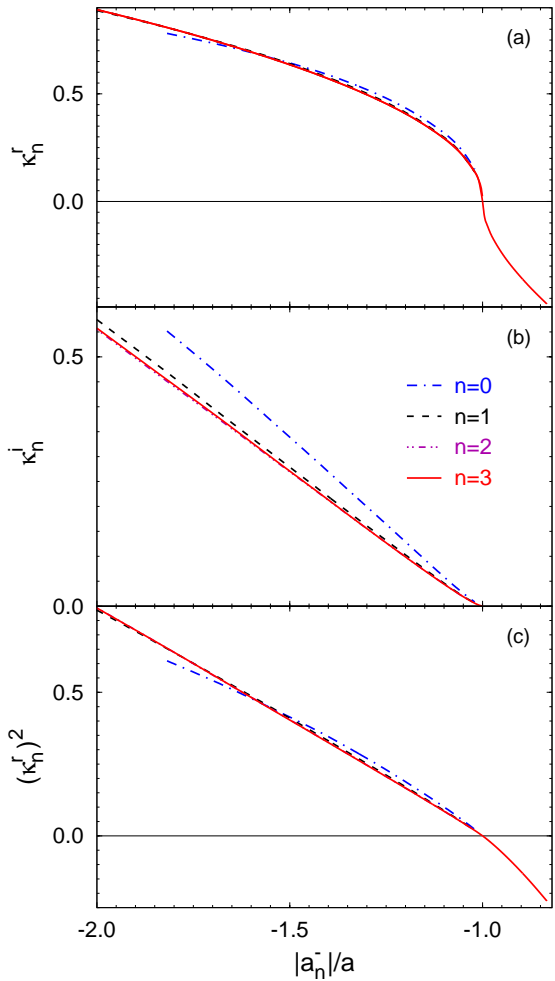}
\end{center}
\caption{\label{fig:k0-3} (Color online)
Components of complex dimensionless resonance momentum
for ground ($n=0$) and excited ($n=1$, 2, and 3) 
three-boson states as functions of the inverse two-boson scattering length.}
\end{figure}

\begin{figure}[!]
\begin{center}
\includegraphics[scale=0.58]{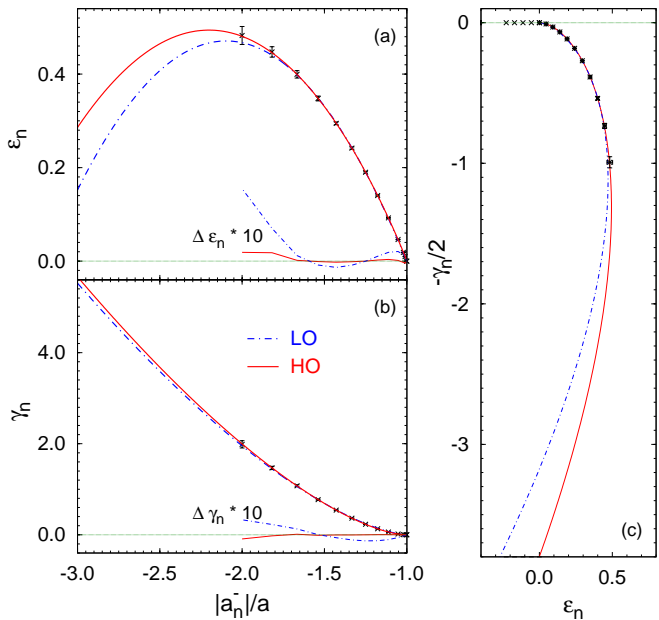}
\end{center}
\caption{\label{fig:egkfit} (Color online)
Dimensionless energy (a) and width (b) for the 
 excited ($n=3$) 
three-boson state as functions of the inverse two-boson scattering length.
The LO (dashed-dotted curves) and HO (solid curves) approximations are compared 
with the results of the direct calculation, represented by symbols with error bars.
Thin curves around the zero line represent the corresponding differences,
amplified by a factor of 10.
The panel (c) shows the resonance trajectory in the complex energy plane.
}
\end{figure}

The interdependence of the two-boson scattering length and 
three-boson energy (or the binding momentum) in the bound state region 
$|\anz|/a > -1$ is well-known and parameterized in Refs.~\cite{naidon:rev,braaten:rev,gattobigio:19a}.
It is desirable to derive  semi-analytical relations also in the resonance region
for the real energy part and width. An insight into a possible form
of the parametrization can be obtained by investigating the dependence
of the dimensionless complex resonant momentum $\kappa^r_n - i \kappa^i_n$,
related to energy and width as
$\varepsilon_n - i \gamma_n/2 = (\kappa^r_n - i \kappa^i_n)^2$,
i.e.,
\begin{subequations} \label{eq:eg-kappa}
\begin{align} 
\varepsilon_n = {} & (\kappa^r_n)^2 - (\kappa^i_n)^2, \\
\gamma_n =  {} & 4 \kappa^r_n \, \kappa^i_n.
\end{align}
\end{subequations}
Note that in the bound-state region 
$\kappa^r_n = k_n |\anz|$ simply relates to the standard binding momentum $k_n$ while
$\kappa^i_n = 0$. The dependence of $\kappa^r_n$ and $\kappa^i_n$
on the inverse scattering length 
is presented in panels (a)-(b) of Fig.~\ref{fig:k0-3}. 
A remarkable feature is that $\kappa^i_n$ depends on 
$|\anz|/a$ almost linearly. Furthermore, $(\kappa^r_n)^2$  displayed in the 
panel (c) of Fig.~\ref{fig:k0-3} also exhibits a nearly linear dependence on
$|\anz|/a$ in the resonance region $|\anz|/a < -1$, while deviations from the linearity are  
more visible in the bound-state region.
Thus,  introducing for brevity the variable $x_a = -(1+|\anz|/a)$, 
 in the resonance region ($x_a > 0$) the behavior of the dimensionless momentum components roughly is 
$\kappa^i_n \sim x_a $ and $(\kappa^r_n)^2 \sim x_a $. Based on these observations
and Eq.~(\ref{eq:eg-kappa}), the
lowest-order (LO) approximate expressions for energy and width are postulated, i.e.,
\begin{subequations} \label{eq:eg-lo}
\begin{align} 
\varepsilon_n \approx {} & c_1^\varepsilon x_a - c_2^\varepsilon x_a^2, \\ 
 \gamma_n \approx {} &  2 c_1^\gamma x_a^{3/2}.
\end{align}
\end{subequations}
The state label $n$ on the right-hand side is suppressed for brevity.
Assuming strictly linear dependence of  $\kappa^i_n$ and $(\kappa^r_n)^2$ on $x_a$,
only two out of three coefficients in Eqs.~(\ref{eq:eg-lo}) would be independent.
However, as there are deviations, the most evident one being for $(\kappa^r_0)^2$,
the quality of the approximation is improved by considering the three
coefficients as independent.
The higher order (HO) approximation is obtained taking into account 
small deviations from linearity in $\kappa^i_n$ and $(\kappa^r_n)^2$ at small $x_a$,
i.e., near the threshold. 
Proposed phenomenological  relations
\begin{subequations} \label{eq:eg-ho}
\begin{align} 
(\kappa^r_n)^2 \approx {} &  c_1^r \, x_a  + c_2^r \, x_a^{2/3}, \\ 
\kappa^i_n \approx {} & x_a ( c_1^i  - c_2^i \, e^{-c_3^i x_a^{1/2}}),
\end{align}
\end{subequations}
are used in Eq.~(\ref{eq:eg-kappa}) for HO approximations of
$\varepsilon_n$ and $\gamma_n$.
The coefficients in Eqs.~(\ref{eq:eg-lo}) and (\ref{eq:eg-ho})  are obtained by
fitting the results presented in Figs.~\ref{fig:Er0-3} - \ref{fig:k0-3}.
In order to check the predictive power of the LO and HO approximations, only
the data points in the regime $-1.67 < |\anz|/a < 1 $ are included in the fit. This way
one can estimate the quality of approximations by comparing them with
data at $|\anz|/a < -1.67$.
For $n=3$, which is expected to be a very accurate approximation of the universal limit,
the values of fit parameters are
\begin{subequations} \label{eq:c-lo}
\begin{align} 
c_1^\varepsilon = {} &  0.8582 \pm  0.0042  , \\
c_2^\varepsilon = {} &  0.3911 \pm  0.0079  , \\ 
c_1^\gamma     = {} &  0.9766 \pm  0.0057  ,
\end{align}
\end{subequations}
and 
\begin{subequations} \label{eq:c-ho}
\begin{align} 
  c_1^r = {} &  0.7416 \pm  0.0018  , \\
  c_2^r = {} &  0.0525 \pm  0.0014  , \\
c_1^i = {} &  0.5671 \pm  0.0009  , \\
c_2^i = {} &  0.4751 \pm  0.0131   , \\
c_3^i = {} &  4.1714 \pm  0.0900
\end{align}
\end{subequations}
The LO and HO approximations for energy 
and width 
 are compared in panels (a) and (b) of Fig.~\ref{fig:egkfit}
with the results of the direct calculation; in all cases $n=3$.
Both LO and HO approximations seem to fit the original data very well
up to $|\anz|/a  > -1.67$. 
However, a closer inspection of deviations
$\Delta \varepsilon_n = \varepsilon_n - \varepsilon_n({\rm XO})$
and $\Delta \gamma_n = \gamma_n -\gamma_n({\rm XO})$, amplified by a 
factor of 10 and shown by thin curves near the zero line, 
demonstrates considerably better accuracy of the HO approximation,
especially for the  two points at the left  that were excluded from the fit.
Nevertheless, the LO still fits those data within their error bars.

For curiosity, in Fig.~\ref{fig:egkfit} the LO and HO results are extrapolated
to the $|\anz|/a < -2$ region. The LO and HO predictions, though quantitatively different,
qualitatively exhibit the same trend, i.e.,
$\gamma_n({\rm XO})$ increases rapidly while $\varepsilon_n({\rm XO})$ reaches its maximum and then decreases
with decreasing $|\anz|/a$, becoming a completely unobservable very broad subthreshold resonance.
The trajectory of the resonance in the complex  plane $\varepsilon_n - i\gamma_n/2$
is shown in the 
panel (c) of Fig.~\ref{fig:egkfit} which uses roughly the same scale for both real
and imaginary part. The shape clearly shows that the resonance does not rise much above the
threshold but broadens rapidly. 
 The lower part of the curves is an extrapolation based on 
Eqs.~(\ref{eq:eg-lo}) and (\ref{eq:eg-ho}),
 but both LO and HO approximations consistently indicate the evolution 
into a subthreshold resonance when $|a|$ decreases.

As mentioned, an earlier study of Efimov resonances was performed by 
Bringas \etal{} \cite{bringas:04a}. They solved bound-state equation with 
contour deformation into the complex plane.
Instead of the inverse two-boson scattering length
$1/a$ they used the square root of the virtual dimer energy, $\sqrt{|B_2|}$.
Since in the universal regime $B_2 = -1/(ma^2)$, these two representations differ by finite-range corrections only, 
allowing for an easy qualitative comparison.
Bringas \etal{} studied $n=0$, 1, and 2 Efimov states, the evolution of their energies and widths is 
qualitatively fully consistent with present results. A perfect agreement between $n=1$ and 2 results 
in Ref.~\cite{bringas:04a} was not achieved,
indicating that the universal regime was not yet reached with a good accuracy. This can be quantified
by ratios of virtual dimer energies $B_2^{(n)} = B_2|_{a=\anz}$ when the $n$-th trimer crosses the threshold.
In detail, $\sqrt{B_2^{(1)}/B_2^{(2)}} = 21.5$ in Ref.~\cite{bringas:04a},
while $\sqrt{B_2^{(1)}/B_2^{(2)}} = 21.74$  and 
$\sqrt{B_2^{(2)}/B_2^{(3)}} = 22.63$ in this work, the universal limit being $22.694$.
Thus, the present results approach the universal limit much closer than those of Ref.~\cite{bringas:04a}.

The dependence of the resonance energy and width on the 
inverse two-boson scattering length was presented also by Jonsell \cite{jonsell:06a}.
However, in that work  deep dimer states were effectively included by using a three-body parameter
with finite imaginary part. This precludes a direct comparison with the present work. Nevertheless,
a very good  qualitative agreement is found between the results in
Figs.~\ref{fig:Er0-3} - \ref{fig:G0-3} and the ones in Fig.~3 of Ref.~\cite{jonsell:06a},
limited to $|\anz|/a > -1.3$, the quantitative differences being of the order of 10\%.
This possibly indicates that deep dimers do not affect significantly the energy and dissociation
width of the Efimov resonance.

Another important and interesting problem is the quantification of finite-range effects.
Since the evolution of the resonance can be reasonably reproduced by a few 
 parameters of the LO approximation (\ref{eq:eg-lo}), they are chosen for this study.
Ji \etal {} showed that finite-range  corrections to various Efimov features in the bound-state regime
can be expressed by terms proportional to  $r_s$ and $n r_s$ \cite{ji:15a}, and related this result to an earlier
work by Kievsky \etal{}  \cite{PhysRevA.87.052719}. The connection between the two approaches was updated
recently by Gattobigio \etal{} \cite{gattobigio:19a} through the running Efimov parameter.
It appears that finite-range effects for the resonance parameters 
$c_j^\varepsilon$ and $c_1^\gamma$ determined in the present work are compatible
with the pattern of Ref.~\cite{ji:15a}. For $n=0$, 1, 2, and 3 
the approximate expressions
\begin{subequations} \label{eq:c-nr}
\begin{align} 
c_1^\varepsilon \approx {} &  0.8581 - 0.4236\, r_s/\anz - 1.1753\, n r_s/\anz  , \\
c_2^\varepsilon \approx {} &  0.3890 - 1.1618\, r_s/\anz - 1.8038\, n r_s/\anz  , \\
c_1^\gamma      \approx {} &  0.9729 - 0.6640\, r_s/\anz - 1.3157\, n r_s/\anz   ,
\end{align}
\end{subequations}
reproduce the results of direct calculations
with good accuracy,  the deviations being below
0.05\%, 1\%, and 0.3\%, respectively.
 However, the above relations involve non-universal
coefficients and cannot predict  finite-range corrections for other force models.

\begin{figure}[!]
\begin{center}
\includegraphics[scale=0.68]{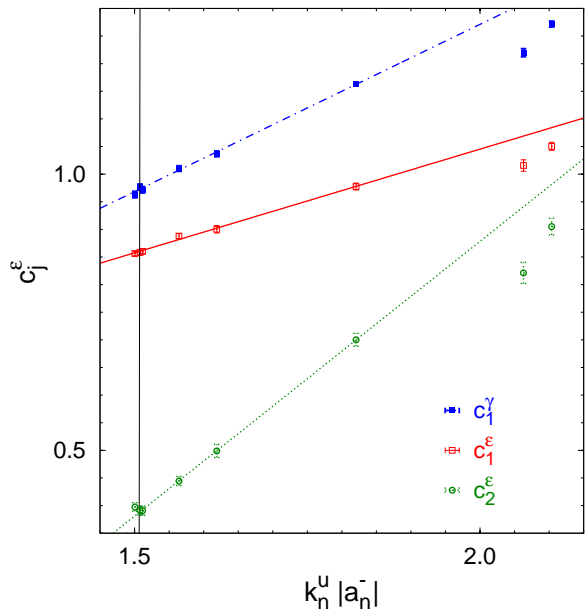}
\end{center}
\caption{\label{fig:c-ka} (Color online)
Resonance energy and width parametrization coefficients $c_j^\varepsilon$ and
$c_1^\gamma$ for various levels and force models as
functions of the product $k_n^u |\anz|$. In terms of the potential (\ref{eq:lsep})
parameter $\beta$ and level $n$ the points from left to right correspond to
$(\beta,n)$ being $(-8,1)$, $(0,3)$, $(0,2)$, $(0,1)$, $(-6,1)$, $(-4,1)$, $(0,0)$,
the last point on the right being the one for the MT potential.
The lines are described by Eqs.~(\ref{eq:c-ka}), while the vertical one
labels the universal limit $k_n^u |\anz| = 1.50763$.
}
\end{figure}

To study finite-range effects, four additional calculations with different force models are performed
for low ($n=0$ or 1) resonances.
One of them uses the scaled Malfliet-Tjon (MT) I-III ${}^1S_0$ potential \cite{tjon:75},
that has an attractive long-range part and a strongly repulsive short-range part. 
However, the results turn out to be quite close to the ones based on the
potential (\ref{eq:lsep}), i.e., $r_s/|\anz| = 0.369$ and $k_n^u |\anz| = 2.1036$ for $n=0$.
Therefore other models are desired to have different
 $r_s/|\anz|$ and $k_n^u |\anz|$ values, spanning the gap between the previously shown $n=0$ and 1 results 
of the potential (\ref{eq:lsep}) with $\beta=0$. This goal can be achieved
by the variation of the parameter $\beta$ in the potential (\ref{eq:lsep}) thereby
enhancing the high-momentum components and enabling  the control
 of $r_s/|\anz|$ and $k_n^u |\anz|$ values. With $n=1$ and $\beta = -4,\ -6$, and $-8$ they become
$r_s/|\anz| = 0.258$, 0.160, and 0.0671,  and  
$k_n^u |\anz| = 1.8206$, 1.6190, and 1.5004, respectively.
Thus, in total this amounts to eight different combinations of force model and level.
The $c_j^\varepsilon$ and $c_1^\gamma$ results  for these eight sets do not exhibit a clear dependence on 
$r_s/|\anz|$ and/or $nr_s/|\anz|$, but show linear correlation with the corresponding
$k_n^u |\anz|$ values as demonstrated  in Fig.~\ref{fig:c-ka}.
Most of the points are very close to the lines parameterized by
\begin{subequations} \label{eq:c-ka}
\begin{align} 
c_1^\varepsilon \approx {} & 0.8604 + 0.3755 (k_n^u |\anz| - 1.50763), \\
c_2^\varepsilon \approx {} & 0.3885 + 0.9949 (k_n^u |\anz| - 1.50763), \\
c_1^\gamma      \approx {} & 0.9730 + 0.6058 (k_n^u |\anz| - 1.50763), 
\end{align}
\end{subequations}
except for the two $n=0$ points possessing largest  $r_s/|\anz|$ and 
$k_n^u |\anz|$ values, for which the finite-range effects are expected to be sizable, thereby
leading to larger deviations.
Nevertheless, Eqs.~(\ref{eq:c-ka}) may be useful for the estimation of the resonance position and width
from the features of the corresponding bound state.


\section{Summary \label{sec:sum}}

The three-boson continuum study was carried out in the regime of negative
two-boson scattering length with no shallow or deep dimers,
where the three-particle Efimov states are not bound but resonant.
Exact scattering equations for three-particle transition operators were solved in the momentum-space framework using
several interaction models. From the energy dependence of the various on-shell and off-shell transition
matrix elements the resonance properties were determined,
in particular, their universal limit  was accurately calculated considering highly excited Efimov states such that
the finite-range effects become negligible. 
Universal relations between the resonance energy and width and the two-particle scattering length $a$ were established,
and simple parametrizations were proposed, based on a nearly linear dependence of the complex resonant momentum components
on the inverse scattering length.
It was found that decreasing $|a|$, i.e., decreasing the attraction,
the resonances rise not much above the threshold, the real part of energy
reaching roughly $\hbar^2/[2m(\anz)^2]$ or $\hbar^2/(8ma^2)$  around $ a \approx \anz/2$,
while the width exceeds $2\hbar^2/[m(\anz)^2]$ at that point, thereby rendering the resonance  practically unobservable.
Extrapolation of the developed parametrizations indicates the evolution into a very broad subthreshold resonance by further
decrease of $|a|$.

In addition, finite-range effects were studied as well,
and their correlation with the bound state properties at unitarity are found.
This may be useful for the estimation of the resonance energy and width using the information from the bound-state regime.

The present work demonstrated the feasibility of accurately extracting the properties of three-boson Efimov resonances,
and thereby encourages the future studies of four-particle Efimov resonances. This constitutes a highly challenging problem
with novel aspects, as the four-boson resonances, due to the presence of lower trimers,
have a finite width already at the threshold.
 The work into this direction is underway.

\vspace{1mm}

The author acknowledges  support  by the Alexander von Humboldt Foundation
under grant no. LTU-1185721-HFST-E.



\end{document}